\documentclass[twoside,11pt]{article}

\usepackage{jmlr2e}
\usepackage{floatrow}
\usepackage{makeidx}  
\usepackage{amsmath}
\usepackage{amssymb}
\usepackage{graphicx}
\usepackage{comment}
\usepackage{color}
\usepackage{soul}
\usepackage{xspace}
\usepackage{caption}
\usepackage{subcaption}
\newfloatcommand{capbtabbox}{table}[][\FBwidth]
\usepackage{blindtext}

\def\Sys{ScriptNet\xspace}
\def\MPL{LaMP\xspace}
\def\SCL{CPoLS\xspace}
\def\Emb{\textsc{Embedding}\xspace}
\def\LSTM{\textsc{LSTM}\xspace}
\def\MaxPool{\textsc{MaxPool1d}\xspace}
\def\relu{\textsc{Relu}\xspace}
\def\RecConv{ \textsc{RecurrentConvolutions}\xspace}
\def\Conv{\textsc{Conv1D}\xspace}

\ShortHeadings{Neural Classification of Malicious Scripts}{Stokes \emph{et al.}}
\firstpageno{1}

\begin{document}

\title{Neural Classification of Malicious Scripts: A study with JavaScript and VBScript}

\author{\name Jack W. Stokes \email jstokes@microsoft.com \\
	\addr Microsoft Research\\
	Redmond, WA 98052, USA
	\AND
	\name Rakshit Agrawal \email ragrawa1@ucsc.edu \\
	\addr Department of Computer Science\\
	University of California, Santa Cruz\\
	Santa Cruz, CA 95064, USA
	\AND
	\name Geoff McDonald \email geofm@microsoft.com \\
	\addr Microsoft Corp.\\
	Vancouver, BC, V6E 4M3, CA}

\editor{}

\maketitle

\begin{abstract}
	Malicious scripts are an
	important computer infection threat vector. Our analysis
	reveals that the two most prevalent types of
	malicious scripts include JavaScript and VBScript.
	The percentage of detected
	JavaScript attacks are on the rise.
	To address these threats, we investigate two deep recurrent
	models, LaMP (LSTM and Max Pooling) and CPoLS (Convoluted
	Partitioning of Long Sequences), which process JavaScript and VBScript as byte sequences.
	Lower layers capture the sequential nature of these byte
	sequences while higher layers classify the resulting
	embedding as malicious or benign.
	Unlike previously proposed solutions, our models are
	trained in an end-to-end fashion allowing discriminative
	training even for the sequential processing layers.
	Evaluating these models on a large corpus of 296,274 JavaScript
	files indicates that the best performing LaMP model has a 65.9\% true
	positive rate (TPR) at a false positive rate (FPR) of 1.0\%. Similarly, the best CPoLS model
	has a TPR of 45.3\% at an FPR of 1.0\%. LaMP and CPoLS yield a
	TPR of 69.3\% and 67.9\%, respectively, at an FPR of 1.0\% on a collection of 240,504 VBScript files.
\end{abstract}

\section{Introduction}
\label{sec:intro}
%
%
Malicious scripts are widely abused by malware authors to infect users' computers. In this paper, we show that in the current threat landscape, the two most prevalent
types of script malware that Windows users encounter are JavaScript (JS) and VBScript (VBS). JavaScript is an interpreted scripting language developed by Netscape that
is often included in webpages to provide additional dynamic functionality~\cite{JS}. VBScript, or Microsoft Visual Basic Scripting Edition, is an active scripting
language originally designed for Internet Explorer and the Microsoft Internet Information Service web server~\cite{VB}.

\begin{figure}[tbh]
	\centering
	{\label{}\includegraphics[width=0.7\columnwidth]{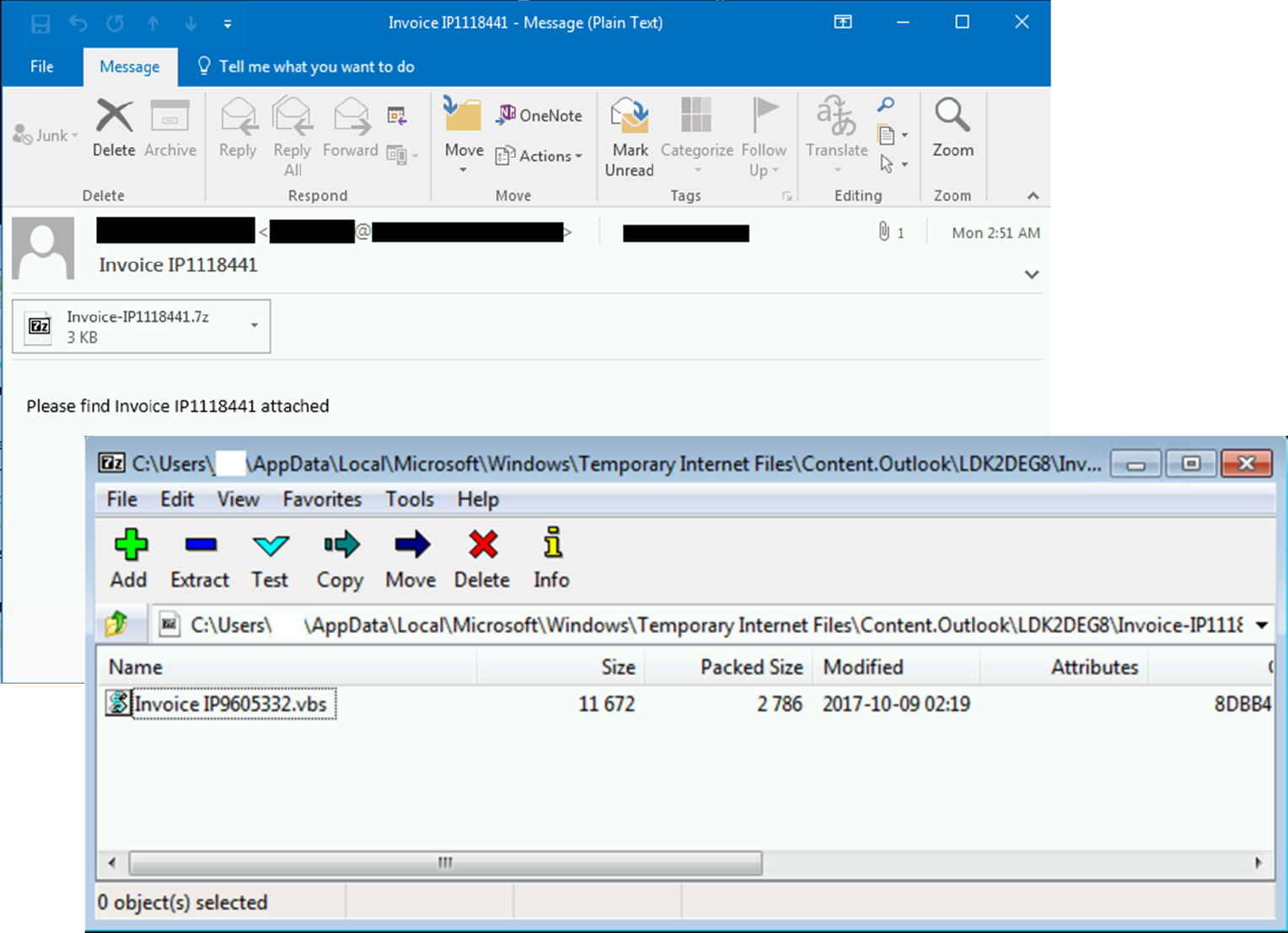}}
	\caption{Example of an email-based social engineering attack using an attached VBScript file.}
	\label{fig:malware_email}
\end{figure}

\begin{figure}[tbh]
	\centering
	\begin{subfigure}{.5\textwidth}
		\centering
		\includegraphics[trim = 2.0in 2.0in 2.0in 2.0in,clip,width=1.0\columnwidth]{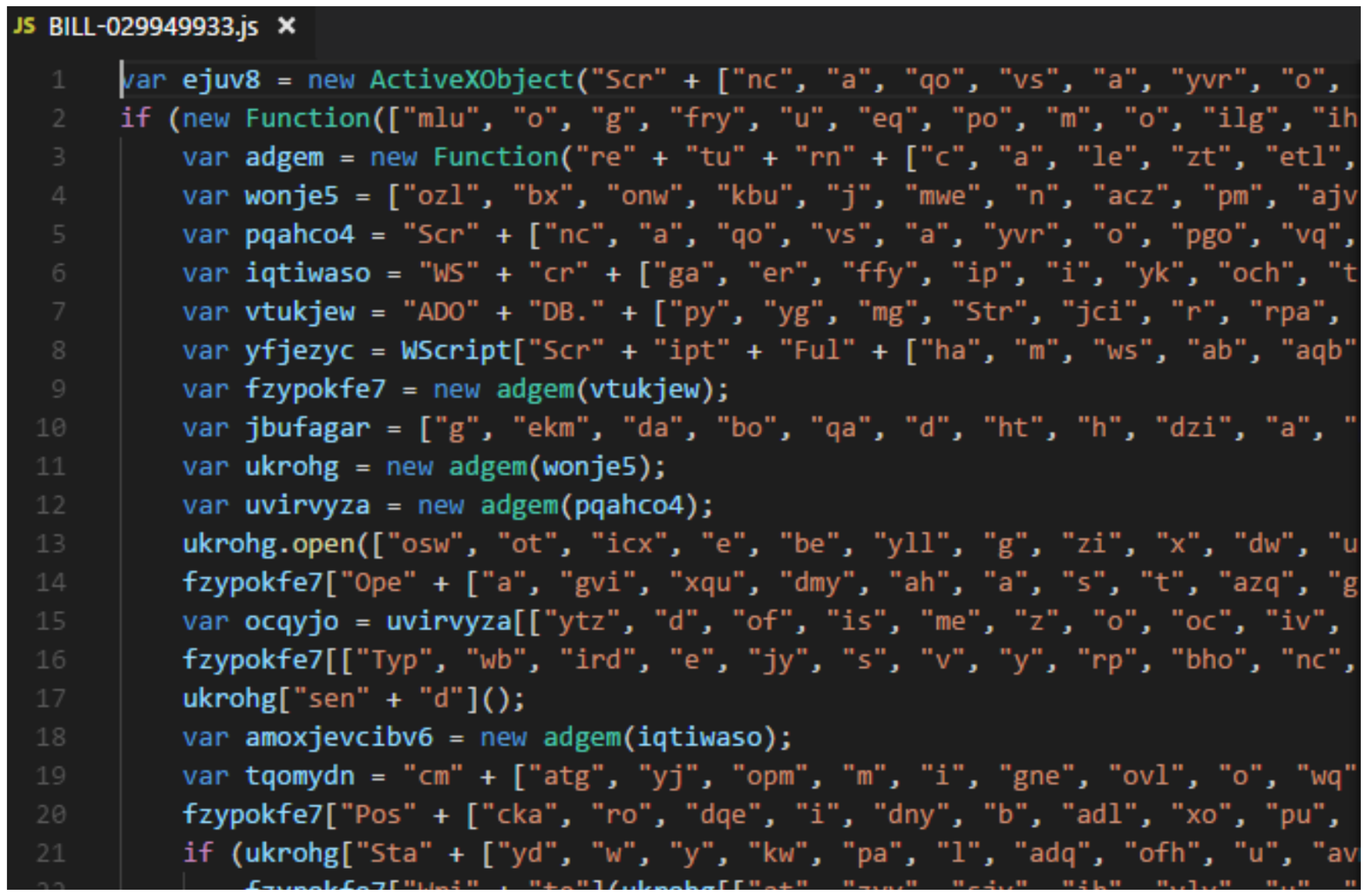}
		\caption{Malicious JavaScript}
		\label{fig:mal-js}
	\end{subfigure}%
	\begin{subfigure}{.5\textwidth}
		\centering
		\includegraphics[trim = 2.0in 2.0in 2.0in 2.0in,clip,width=1.0\columnwidth]{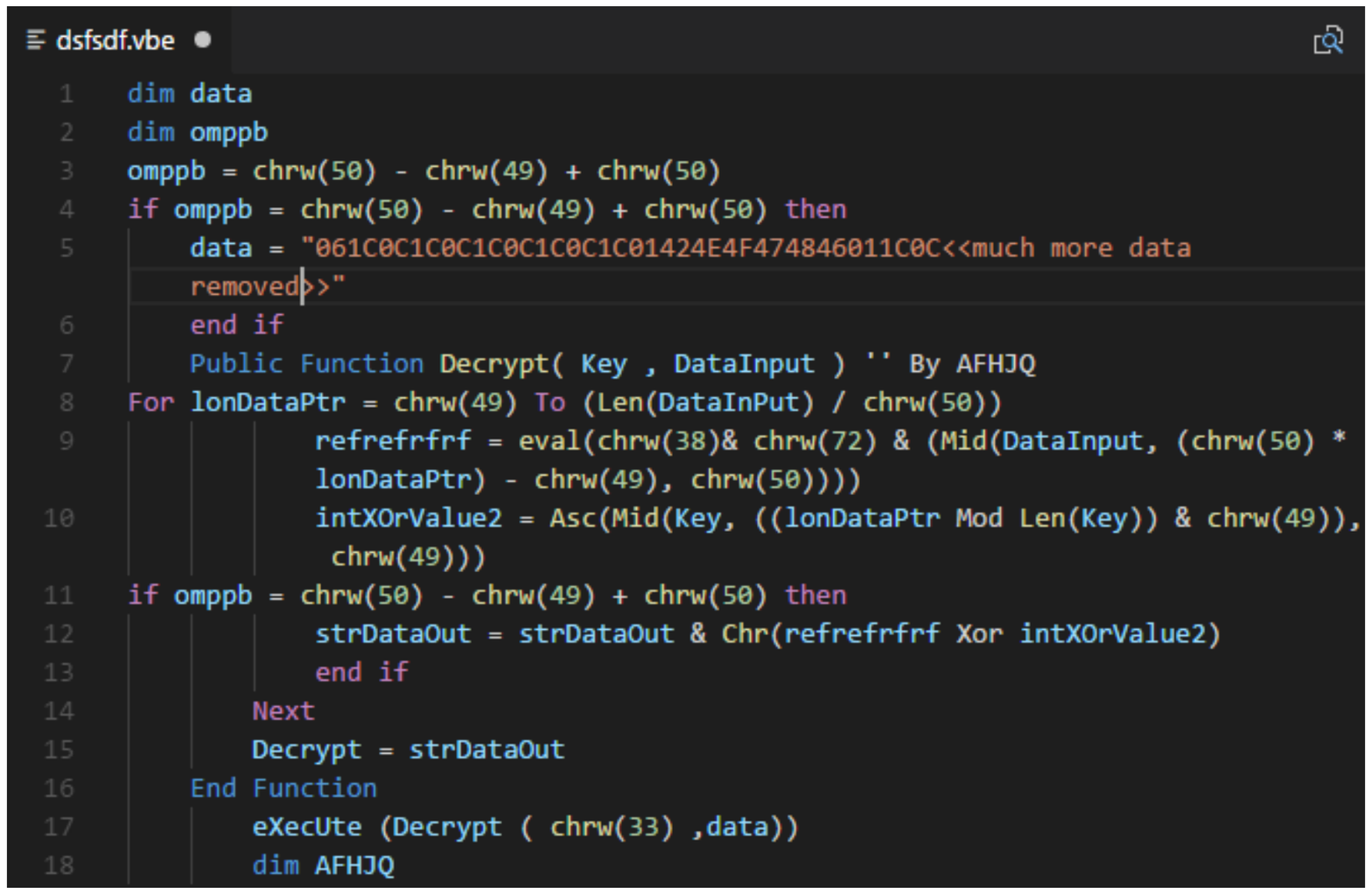}
		\caption{Malicious VBScript}
		\label{fig:mal-vbs}
	\end{subfigure}
	\caption{Example a) JavaScript file from the TrojanDownloader:JS/Swabfex malware family, and b) from a malicious VBScript file from the Worm:VBS/Jenxcus malware family.}
	\label{fig:mal-script}
\end{figure}

Spearphishing attacks have been a key component of several recent large-scale data breaches~(\cite{crnBreach,verizonBreach}).
For example in Figure~\ref{fig:malware_email}, a typical spearphishing attack involves a user being sent an email stating that they have an outstanding invoice. An archive is attached to the email, and inside the archive is a VBScript file called "invoice.vbs". If the user opens the VBScript file, it will be executed through the default file association using a native script execution host on Windows (in this example ``wscript.exe''). Now that the malicious script is running on the computer, these attacks commonly download and execute further malware such as ransomware~(\cite{mmpc_2016_js_ransomware}). Figure~\ref{fig:mal-script} presents examples of malicious JavaScript and VBScript content.

While a wide range of different machine learning models have been proposed for detecting
malicious executable files~(\cite{Gandotra2014}), there has been little work in investigating malicious JavaScript, and even less
research has been devoted to trying to detect malicious VBScript. Previous JavaScript solutions include those based on static analysis~(\cite{Likarish2009,Maiorca15,Shah2016}),
and both static and dynamic analysis~(\cite{Corona2014}). Two previous solutions for VBScript are based on static analysis~(\cite{Kim2006,Wael2017}).
In addition, deep recurrent models have recently been proposed detecting system API calls in PE files~(\cite{BenMalware,Kolosnjaji,PascanuMalware}), JavaScript~(\cite{Wang2016}),
and Powershell~(\cite{Hendler2018}).

There are several challenges posed by trying to detect malicious JavaScript and VBScript. One main challenge is the lack of labeled data. While
obtaining malicious samples is challenging enough, creating a large benign set of script files is extremely difficult given strict privacy
email policies which prevent manual inspection of undetected email.
Furthermore, malicious scripts include obfuscation to hide the malicious content, and often unpack or decrypt the underlying malicious script only upon execution. Complicating this is the fact that the obfuscators, in some cases, are used by both benign and malware files. Thus pure static analysis of the primary script often fails to detect some malicious activity. Another problem is that anti-virus (AV) automation systems such as sandboxing environments are designed primarily to handle Windows Portable Executable (PE) files (\textit{e.g.}, .exe and .dll).
%
Accordingly, the number of labeled script files is typically much lower than for executable files.

In this paper, we propose \Sys, a deep recurrent neural classification system which can be trained to detect either malicious JavaScript or VBScript using a combination of both
static and dynamic analysis. We first use a production anti-virus engine to dynamically execute
a script in a sandboxed environment inside of the engine. This
allows the AV engine to safely analyze any child scripts which are dropped during script execution without infecting the
computer.

We investigate two different models for the task of detecting malicious JavaScript and VBScript.
Both models encode sequential information using one or more long, short-term memory (LSTM) layers.
The LSTM and Max Pooling (\MPL) model follows a two-stage approach where the first stage learns a language model
for the individual characters in the script content. Next, the second stage includes a, potentially deep, neural network
for the final classification of the script as malicious or benign.
To allow the processing of longer script files, we next investigate the Convoluted Partitioning of Long Sequences (\SCL)
model which adds an additional layer consisting of a one-dimensional convolutional neural network.
\MPL is similar to the model
proposed by~\cite{BenMalware} for PE files, but differs in two respects. While Athiwaratkun's model
also has an LSTM-based language model followed by a neural network classification stage, each component is trained
in isolation. The language model is first trained in an unsupervised fashion, and this trained language model
is then frozen and used to generate the embeddings for the classification stage. Instead, \MPL is trained
with end-to-end learning where all the model parameters, including those in the language model and the classifier,
are learned simultaneously directly from the characters in the script content. Similarly, \SCL is also trained
in an end-to-end manner. Second, \MPL extends the model in~\cite{BenMalware} to allow for stacked (\textit{i.e.}, multiple) LSTM
layers.
Since our models operate directly on the script content encoded as bytes, they do
not require careful and potentially computationally
expensive feature engineering proposed by other solutions.
The main contributions of this paper include:
\begin{itemize}
	\item{ We study the detection percentage and threat vectors of malicious JavaScript and VBScript from telemetry generated by a production
		anti-virus product. }
	\item{ We investigate two deep recurrent neural network models for the detection of malicious JavaScript and VBScript.}
	\item{ We evaluate these models on two large corpora of JavaScript and VBScript files.}
\end{itemize}




\begin{figure}[tb]
	\begin{floatrow}
		\ffigbox{%
			\includegraphics[width=0.9\columnwidth]{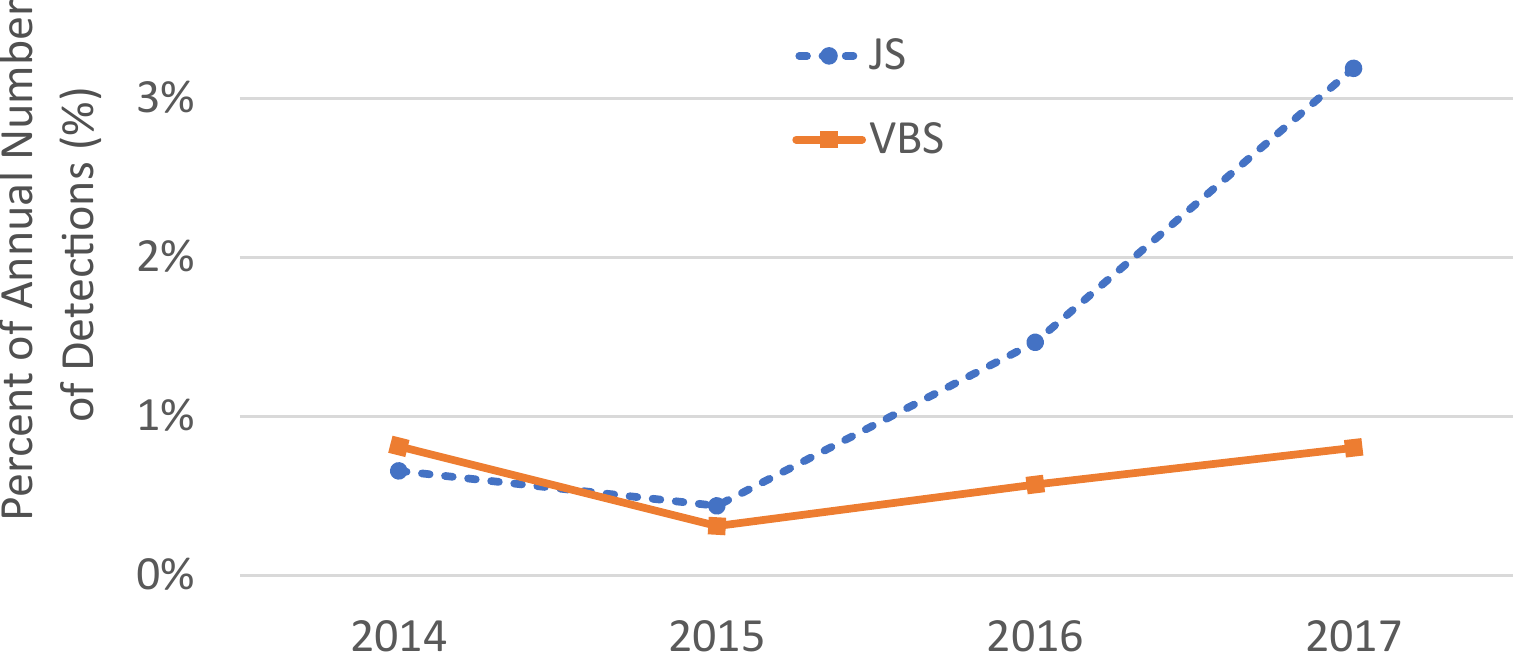}
		}{%
			\caption{Percentage of malicious files detected by the Windows Defender anti-malware engine over time in the categories of JavaScript and VBScript attacks.}%
			\label{fig:malware_trends}
		}
		\ffigbox{%
			\includegraphics[width=0.9\columnwidth]{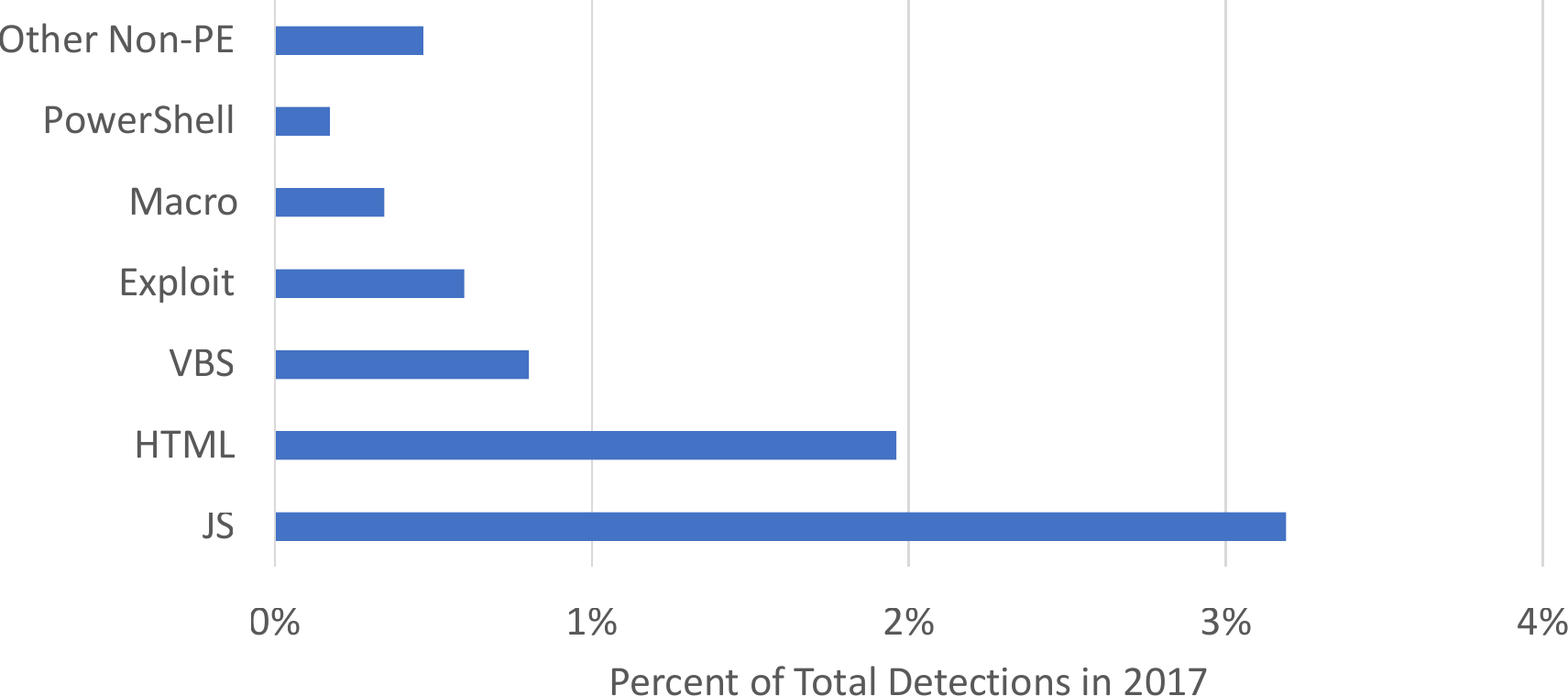}
		}{%
			\caption{Percentage of total detections by file type in 2017. The remaining 92.5\% of detections were for PE files.}
			\label{fig:malware_types}
		}
	\end{floatrow}
\end{figure}

\section{Motivation}
\label{sec:mot}
The detection of malicious JavaScript and VBScript is important for protecting users against modern malware attacks. With advances in browser and operating system security making browser exploit attacks more difficult, miscreants are instead relying on social engineering attacks. Figure~\ref{fig:malware_trends} illustrates the percentage of malicious detected files by the Windows Defender anti-malware engine in the categories of JavaScript and VBScript attacks. The percentage of malicious JavaScript-based attacks has been rising recently, while the percentage of detected attacks involving VBScript have remained relatively constant since 2014.
Figure~\ref{fig:malware_types} indicates the percentage of all, non-PE files detected in the Windows Defender telemetry. This
figure indicates that JavaScript and VBScript are the two most prevalent types of detected scripts found in the telemetry data. Since the remaining 92.5\% of the detections are for PE files, malicious scripts are still a small minority of the detected files in the wild.

\begin{figure}[tbh]
	\centering
	{\label{}\includegraphics[width=0.9\columnwidth]{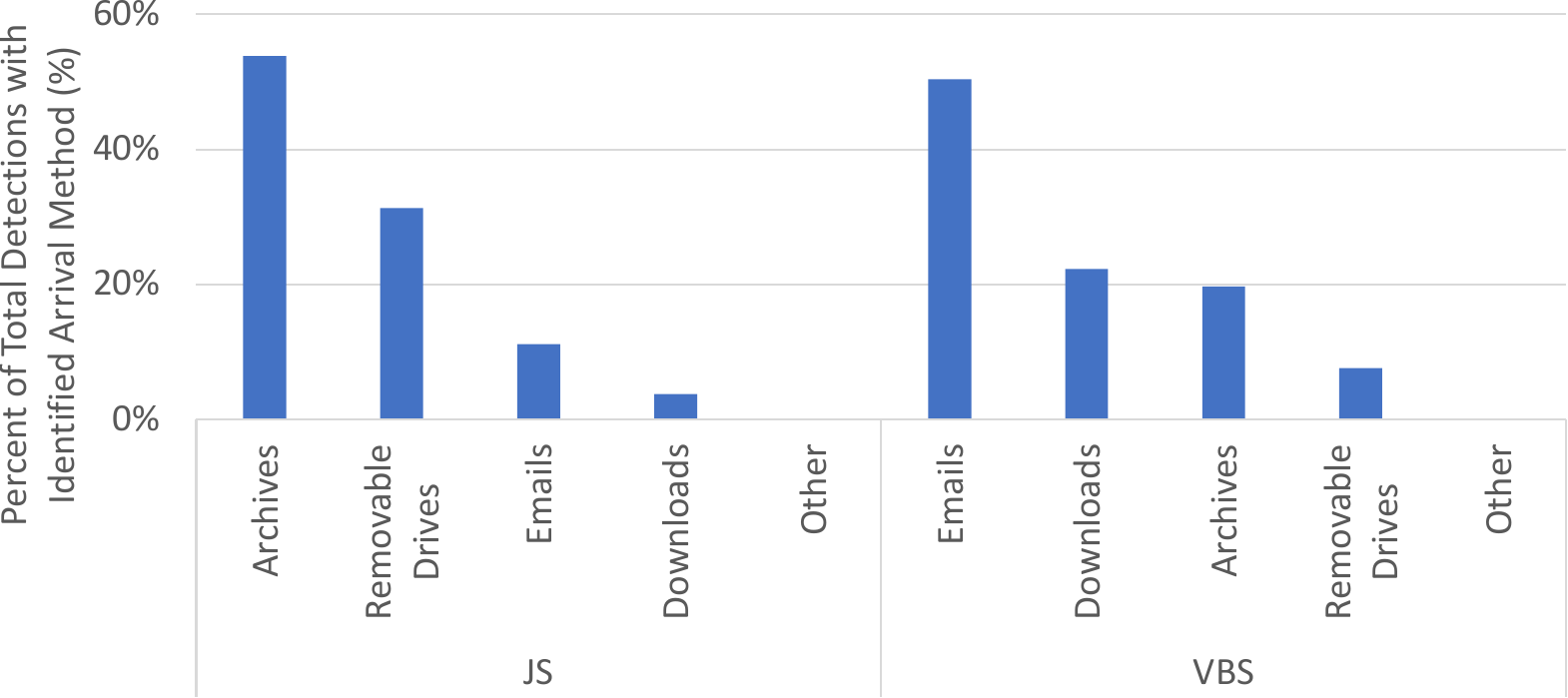}}
	\caption{Arrival methods for malicious JavaScript and VBScript files detected by the Windows Defender anti-malware engine in 2017.}
	\label{fig:malware_vectors}
\end{figure}

Based on the identified arrival methods of malicious JavaScript and VBScript, Figure~\ref{fig:malware_vectors} illustrates the identified attack methods based
on the telemetry data from 2017. Archive file detections, the most prevalent threat vector for JavaScript, are generated when the user extracts the script from within an archive and are often used in social-engineering attacks.
Interestingly, removable drives (\textit{e.g.}, thumbdrives, external USB harddrives) were responsible
for the second most JavaScript attacks.
Only 11.1\% of detected malicious JavaScript files were encountered from malicious email, and 3.8\%
of the files were directly downloaded from the internet.

The distribution of the attack sources for malicious VBScript tells a different story. The main threat vector of malicious VBScript
is emails followed closely again by downloads. Archives and removable drives play a smaller role in VBScript attacks, but they are still important threat vectors.

\section{Threat Model}
\label{sec:threat}
It is necessary to specify the assumptions that we make about the attacker.
The most important assumption is that the model is able to learn some
deep embedding which is able to identify activity related to malware
from the first $T$ bytes (\textit{e.g.}, 200, 1000) of the script.
If the first $T$ bytes are randomly initialized, the models will fail
to detect the activity that somehow captures malicious intent.

Another assumption is that the behavior which identifies an unknown malicious
script is also found in labeled scripts in the training set.
If the training set does not contain scripts which are somehow
related to the unknown script being evaluated, the classifier
may again fail to accurately predict the script type.

As part of the scanning process, the anti-malware engine emulates
an unknown file and attempts to extract any child scripts. It
may be possible that the anti-malware engine fails to
successfully extract all the child scripts. In this case,
the model may also fail to detect the malicious script if the parent
script is predicted to be benign, and the child script which executes the malicious activity is not successfully
extracted.
%

\section{Data}
\label{sec:data}
\textbf{Scripts:}
Building a dataset of malicious and benign scripts for training is a challenge. A sizable percentage of
malicious scripts are delivered in email and for privacy reasons cannot be collected. For this research,
samples were selected randomly from the files observed on users' computers during June 2017 that
had been successfully collected, with permission, by the Windows Defender backend. These samples
are collected by many sources including users directly submitting suspicious files for analysis,
files shared through sample exchanges such as VirusTotal, and scripts that were extracted
from installer packages or archives.




\noindent
\textbf{Labels:} Another challenge in training a classifier for detecting malicious scripts is obtaining
enough labeled data. Since we are trying to predict if a script is malware or benign, we must
obtain both types of labels.

A script is labeled as malware if it has been inspected by our AV partner's analysts and determined to be malicious. In addition, the script is labeled as malicious if it has been detected by the company's detection signatures.
Finally, scripts are
labeled as malware if eight or more other anti-virus vendors detect the script as malware.

Obtaining enough benign scripts is a challenge because labeling a script as benign often requires manual inspection.
Thus, a script is labeled as benign by a number of methods.
First, the script is considered benign if it has been labeled as benign by an analyst or has been collected by a trusted source
such as being downloaded from a legitimate webpage. However, this does not provide enough labeled benign scripts so we augment
this benign dataset with scripts which are not detected by any trusted scanner at least 15 days after our AV partner has first encountered it in the wild.

\noindent
\textbf{Datasets:} Our anti-virus partners provided the first 1000 bytes of 296,274 JavaScript files which contained 166,179 malicious and 130,095 benign
scripts. We randomly assigned these scripts into training, validation, and test sets containing 207,392, 29,627, and 59,255 samples, respectively.  The validation set is a small dataset which is used for
hyperparameter tuning during the training phase.  By doing so, we are later able to make a fair assessment of the
final model's performance on the held-out test set.
Similarly, our partners provided a VBScript dataset with 240,504 examples including 66,028 malicious scripts and 174,476 benign
scripts. This dataset was then randomly split into 168,353 training scripts, 24,050 validation
scripts, and 48,101 test scripts.

\section{System}
\label{sec:system}
\begin{figure}[tbh]
	\centering
	{\label{}\includegraphics[trim = 1.0in 2.0in 2.0in 3.0in,clip,width=0.9\columnwidth]{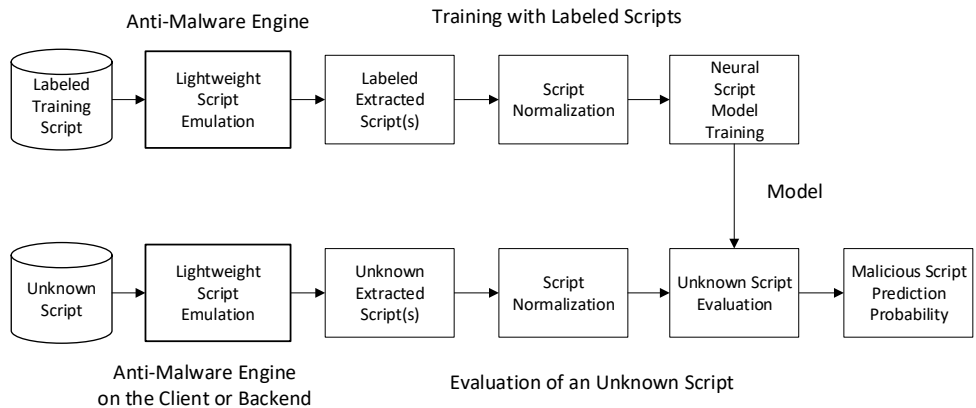}}
	\caption{Overview of the neural script classification system.}
	\label{fig:system}
\end{figure}

Figure~\ref{fig:system} presents an overview of the proposed neural script classification system.
The labeled collection of malicious and benign scripts (\textit{e.g.}, JavaScript or VBScript files), described in the previous section,
are first scanned with the Windows Defender anti-malware engine. During this
scanning operation, the script is emulated and unpacked and may drop one or more additional scripts.
Each child script is also emulated and unpacked which may generate even more scripts. This process continues
until all scripts have been extracted and scanned.

\begin{figure}[tbh]
	\centering
	{\label{}\includegraphics[trim = 0.0in 0.0in 0.0in 0.25in,clip,width=0.95\columnwidth]{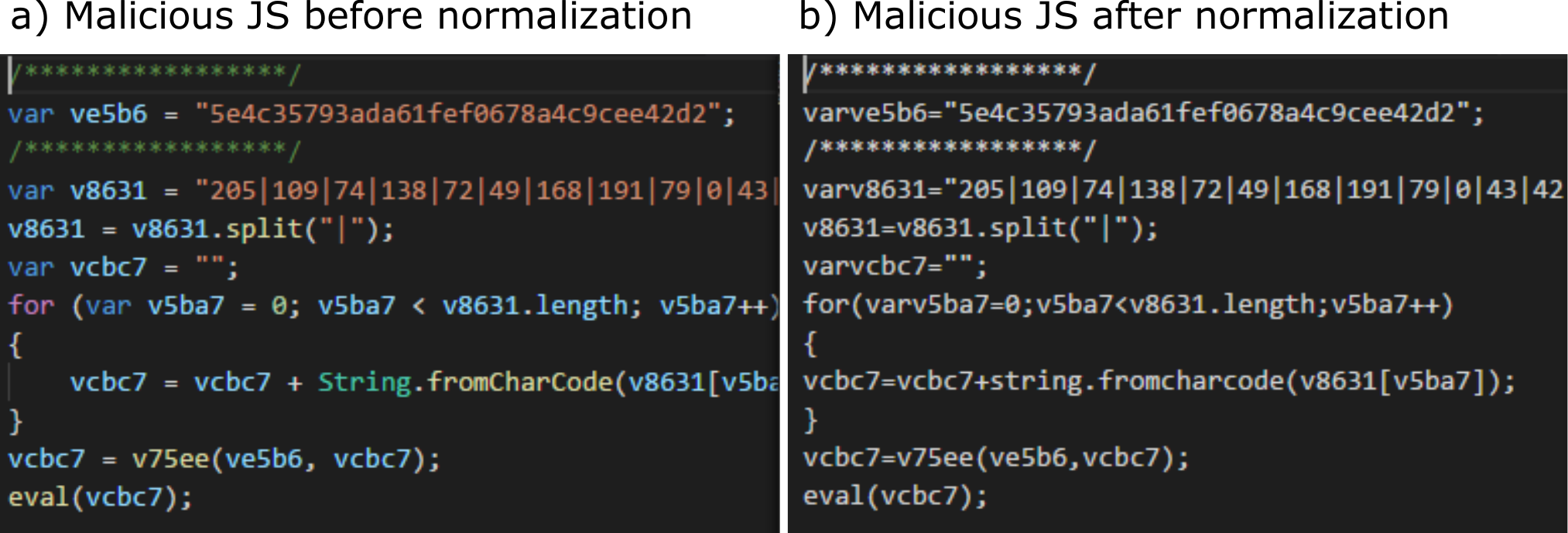}}
	\caption{Example malicious packed JavaScript file from the \textit{TrojanDownloader:JS/Crimace.A} malware family before (left), and after normalization (right).}
	\label{fig:js_before_after_normalization}
\end{figure}

These scripts are next normalized. All whitespace characters, except line breaks, are first removed. Next the text is standardized to lowercase and converted to the US-ASCII character set. Any characters which are not included in the US-ASCII character set, such as non-English language characters, are replaced by the constant character `?'. Figure~\ref{fig:js_before_after_normalization} illustrates an example script before and after normalization.

Before training the model, each normalized script is written to the file system.
To avoid storing malicious content on the hard drive, the characters are next encoded by their
numeric ASCII encoding (\textit{e.g.}, '97' for the character 'a')
delimited by commas. This delimited, encoded sequence data is then used to train the neural script malware model.

To evaluate an unknown file, the system uses the trained model to produce a prediction which indicates the probability that the unknown script is malicious.

\section{Models}
\label{sec:model}
Static and dynamic analysis of script files, like VBScript and JavaScript, allows our system to use
information hidden
in the script's unpacked content to learn its malicious nature.
In this section, we discuss our models which can capture the script files and learn malicious intent using
neural classifier models and sequential learning.

\noindent
\textbf{Translation to Sequences:}
The raw scripts can be considered to be documents containing a limited vocabulary set.
As such, the scripts are long ordered sequences of encoded characters.
For normalized script files, we define our vocabulary as the set of all possible bytes (8-bits).
This leads to a vocabulary of size 256.
Each normalized script, therefore, is a sequence of these bytes.

\noindent
\textbf{Sequential Learning:}
In language models over document-like datasets, sequential learning is a commonly used learning methodology~(\cite{Jozefowicz,SutskeverSeq2Seq}).
Neural network-based models for sequential learning use Recurrent Neural Networks (RNNs), and their variants, to capture the ordered nature of elements, while learning generally over each individual item.
In our models, we use a specific memory-based gated variant of RNNs, known as the Long Short-Term Memory (LSTM)
model~(\cite{GersJj1999,Hochreiter1997}).
LSTMs are used extensively for processing long sequences of data.
In speech and language models in particular, enhanced LSTMs define the state-of-the-art~(\cite{Cho2014,GravesSpeech,Graves2013b,SutskeverSeq2Seq}).
However, their general neural nature, along with the ability to learn using backpropagation through time~(\cite{Werbos1990}),
makes them useful in many domains.
For our byte sequences, we therefore use LSTMs as the primary element for the capturing sequential attributes of the data.
LSTMs can often be implemented with minor variations in their structure.
The implementation used in our models, at each timestep $t$, is described by the following equations:

\begin{equation}
\begin{split}
& \mathbf i_t = \sigma(\mathbf W_{hi} * \mathbf h_{t-1} + \mathbf  W_{xi} * x_t + \mathbf b_i)\\
& \mathbf f_t = \sigma(\mathbf  W_{hf} * \mathbf h_{t-1} + \mathbf W_{xf} * x_t + \mathbf b_f) \\
& \mathbf o_t = \sigma(\mathbf W_{ho} * \mathbf h_{t-1} + \mathbf W_{xo} * x_t + \mathbf b_o) \\
& \mathbf c_t = \mathbf f_t \odot \mathbf c_{t-1} + \mathbf i_t \odot \tanh(\mathbf  W_{hc} * \mathbf h_{t-1} + \mathbf
W_{xc} * x_t + \mathbf b_c) \\
& \mathbf h_t = \mathbf o_t \odot \tanh(\mathbf c_t)
\end{split}
\end{equation}

\noindent where the nonlinearity defined by $\sigma$ corresponds to the logistic sigmoid function. The variables
$\mathbf i_t, \mathbf f_t, \mathbf o_t, \mathbf c_t$ are the input gate, forget gate, output gate and
cell activation, respectively. $\mathbf W_{h*}(\cdot)$ are the weight matrices for each gate corresponding
to the recurrent input from the previous timestep, $\mathbf W_{x*}(\cdot)$ are the input weight matrices per gate,
and $\mathbf b_{*}(\cdot)$ are the biases for each gate.
The function $\odot$ represents the pairwise product between two vectors.

The network takes input vector $x_t$ at each timestep $t$, and updates two properties of the LSTM.
It updates the cell memory $\mathbf c_t$ using the gates as well as the cell memory $\mathbf c_{t-1}$ from the previous timestep.
It then updates the hidden activation $\mathbf h_t$ for timestep $t$ by using the gates and cell memory.
The input vector provided to the LSTM cell can be of any structure depending on the data.
In a categorical representation, it can be a one-hot encoded vector, while in the case of embeddings, it can
be in the form of a dense vector. For sparse featured data, the input can simply be a sparse vector.

\noindent
\textbf{Model Architectures:}
In our experiments for sequential learning, we designed two neural model architectures.
The primary difference in these two architectures is their
resilience against very long length sequences. We will discuss these properties in detail below.

\begin{figure*}[tbh]
	\centering
	{\label{}\includegraphics[trim = 0.0in 3.0in 0.0in 1.0in,clip,width=0.9\columnwidth]{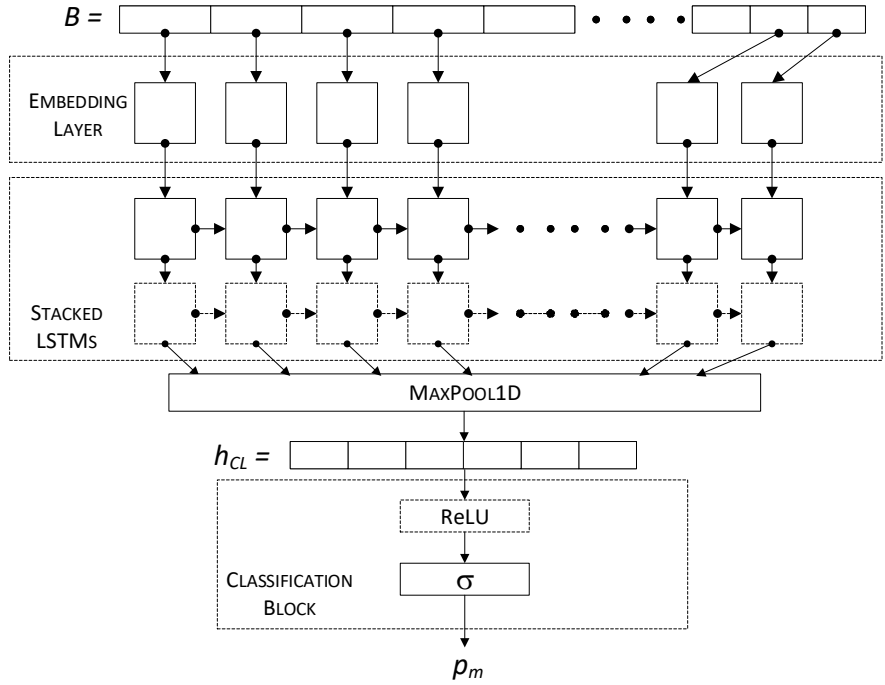}}
	\caption{LaMP model for detecting malicious JavaScript and VBScript files.}
	\label{fig:model}
\end{figure*}

\noindent
\textit{LSTM and Max Pooling:}
In the LSTM and Max Pooling (\MPL) architecture,
illustrated in Figure~\ref{fig:model}, we first use an embedding layer, \Emb, to process the input byte sequence $B$.
Since each element in $B$ corresponds to a byte from the vocabulary, it is symbolic in nature.
We use the embedding layer to transform each byte into a dense vector (\textit{i.e.}, an embedding) which captures relatedness among
different bytes, thereby assisting the overall model in learning.
The sequence of embeddings $E$ is then passed through multiple \LSTM layers stacked
on top of each other.
The \LSTM generates representations for each element in the input sequence as $H_L$.
In order for us to perform classification on the sequence and identify its hidden malicious content,
we transform the sequence $H_L$ into a vector highlighting significant information, while reducing its dimensionality.
For this purpose, we use a temporal, max pooling layer, \MaxPool, as proposed by~\cite{PascanuMalware}.
Given an input vector sequence  $S = [s_0, s_2, \ldots s_{M-1}] \in S$ of length $M$,
where each vector $s_i \in \mathbb{R}^k$ is a $k$-dimensional vector, \MaxPool computes
an output vector $s_{MP} \in  \mathbb{R}^k$ as
$s_{MP}(k) = \max(s_0(k), s_1(k), \cdots s_{M-1}(k))$.

We pass the sequence $H_L$ through \MaxPool to obtain vector $h_L$.
Next, $h_L$ is passed through one or more dense neural layers employing a rectified linear (\relu) nonlinear activation function.
This helps learn an additional layer of weights before performing the final prediction.
The \relu activated vector is finally used by a sigmoid layer to generate final
probability $p_m$ indicating if the script is malicious or benign.
We can formally define $\textsc{\MPL}$ on an input byte sequence $B$ as:
\begin{equation}
\begin{split}
& E = \Emb(B)\\
& H_L = \LSTM(E) \\
& h_L = \MaxPool(H_L) \\
& h_{CL} = \relu(W_L * h_L) \\
& \mathbf{p_m} = \sigma(W_D * h_{CL})
\end{split}
\end{equation}
\noindent where $W_L$ is the weight matrix for the dense \relu hidden layer, and $W_D$ is the weight matrix
for the final sigmoid classification layer.

While \MPL provides a simple model to capture sequences directly, it is limited by the length of the input sequences.
As the length of input sequence $B$ increases, the model becomes both difficult to train and more memory-intensive.
In the case of detecting malicious content, long sequences can often separate two or more bytes far from each other
even when their combined presence is a cause of the malicious intent. When learning directly on a sequence,
it is possible for the model to lose the context of an identified byte earlier in the sequence when processing a new byte
at a larger distance.
To cope with such problems in detection, we therefore, propose another architecture called
Convoluted Partitioning of Long Sequences (\SCL).

\begin{figure*}[tbh]
	\centering
	{\label{}\includegraphics[trim = 0.0in 0.5in 0.0in 0.5in,clip,width=0.7\columnwidth]{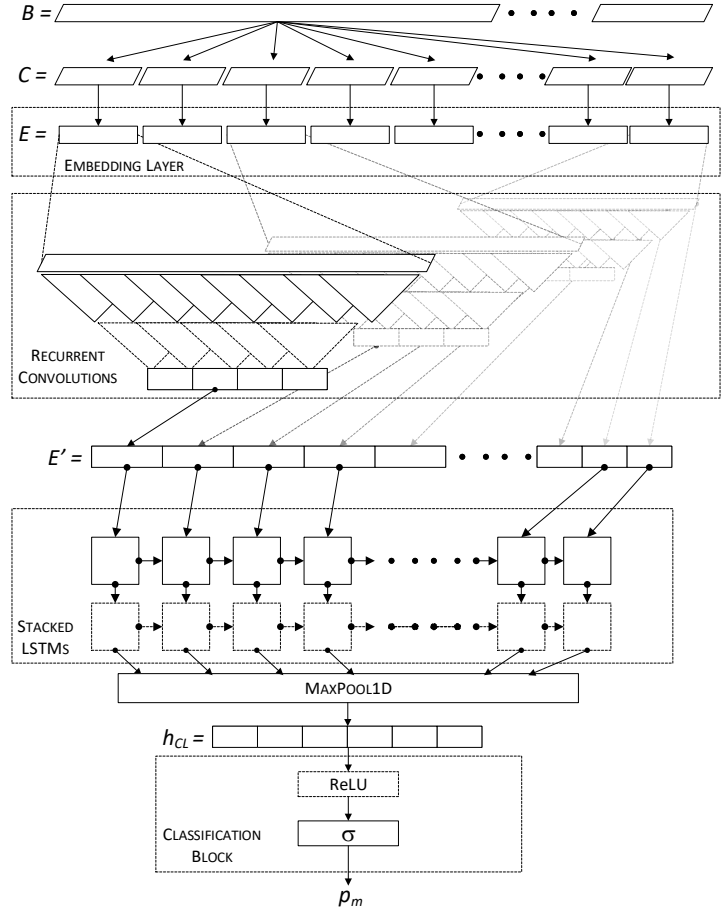}}
	\caption{Convoluted Partitioning of Long Sequences (\SCL) model for detecting malicious JavaScript and VBScript files.}
	\label{fig:scl}
\end{figure*}

\noindent
\textit{Convoluted Partitioning of Long Sequences:}
Convoluted Partitioning of Long Sequences (\SCL) is a neural model architecture designed specifically
to extract classification information hidden deep within long sequences.
In this model illustrated in Figure~\ref{fig:scl}, we process the input sequence in parts by splitting it first into smaller pieces of fixed length.
By performing this step, we generate a sequence of multiple partitions, each of which is a sequence in itself of a smaller length.

We use Convolutional Neural Networks (CNNs)~\cite{LeCun1995ConvolutionalNF} in this model, along with the other \MPL modules.
CNNs are widely used in computer vision~(\cite{krizhevsky2012imagenet,russakovsky2015imagenet}), and
they have also recently shown success in sequential learning domains as well~(\cite{Gehring2016,Gehring2017}).

Given an input byte sequence $B$, the model first splits it into
a partitioned list $C$ containing several small subsequences $c_i \in C$ where $i$ is the index of each partition in $C$.
To translate the bytes in these sequences from symbols to dense vectors, we pass them through
an embedding layer, \Emb, and obtain sequence $E$, where each element $e_i \in E$ corresponds to the
sequence of embeddings for partition $c_i$ in $C$.
Each of these partitions $e_i$, are now separately processed, while still maintaining their overall sequential nature.
We call this method \RecConv.
In this method, we pass each partition $e_i$ through the one-dimensional CNN, \Conv, which applies multiple filters on
the input sequence and generates tensor $e^\chi_i$  representing the convoluted output of vector sequence $e_i$.
$\chi$ refers to the sequence with \Conv performed on it.
The combined list of these convolved partitions $e^\chi_i$ is referred to as $E^\chi$.
In \RecConv, we then reduce the dimensionality of $e^\chi_i$  by performing a temporal max pooling \MaxPool.
\MaxPool takes a tensor input $e^\chi_i$ and extracts a vector $e'_i$ from it.
Similarly, we apply \RecConv on each partition $e_i$ to obtain the updated vectors $e'_i$.
These vectors $e'_i$ are finally combined in the same order to create an updated sequence $E'$ of learned partition representations.
With the help of partitioning, the length of $E'$ is also limited to a trainable length.
%

At this stage, the model uses sequence $E'$ as an input to the \MPL model and learns the probability $p_m$ that the script is malicious.
Therefore, we use a combination of an \LSTM, a second \MaxPool layer, dense \relu activations,
and a final sigmoid layer for generating the prediction $p_m$ on the new input sequence $E'$.
Formally, we define the \textsc{\SCL} model as:
\begin{equation}
\begin{split}
& C = \textsc{Partition}(B)\\
& E = [\Emb(c_i) \hspace{10pt}\forall c_i \in C]\\
& E^\chi = [\Conv(e_i) \hspace{10pt} \forall e_i \in E]\\
& E' = [\MaxPool(e^\chi_i) \hspace{10pt} \forall e^\chi_i \in E^\chi]\\
& p_m = \MPL(E')
\end{split}
\end{equation}
Such a model is resilient to extremely long sequence lengths and can also find malicious objects hidden very late in the
sequence.

\begin{table*}[tb]
	\begin{center}
		\begin{scriptsize}
			\begin{tabular}{| c | c | c | c | c |}
				\hline
				Script Type & Model & Parameter & Description & Value \\
				\hline
				\hline
				JavaScript & LaMP & $B_{JS,LaMP}$  & Minibatch Size & 200 \\
				JavaScript & LaMP & $H_{JS,LaMP}$  & LSTM Hidden Layer Size & 1500 \\
				JavaScript & LaMP & $E_{JS,LaMP}$  & Embedding Layer Size & 64 \\
				\hline
				JavaScript & CPoLS & $B_{JS,CPoLS}$  & Minibatch Size & 50 \\
				JavaScript & CPoLS & $H_{JS,CPoLS}$  & LSTM Hidden Layer Size & 1500 \\
				JavaScript & CPoLS & $E_{JS,CPoLS}$  & Embedding Layer Size & 64 \\
				JavaScript & CPoLS & $W_{JS,CPoLS}$  & CNN Window Size & 10 \\
				JavaScript & CPoLS & $S_{JS,CPoLS}$  & CNN Window Stride & 5 \\
				JavaScript & CPoLS & $F_{JS,CPoLS}$  & Number of CNN Filters & 128 \\
				\hline
				VBScript & LaMP & $B_{VBS,LaMP}$  & Minibatch Size & 100 \\
				VBScript & LaMP & $H_{VBS,LaMP}$  & LSTM Hidden Layer Size & 1500 \\
				VBScript & LaMP & $E_{VBS,LaMP}$  & Embedding Layer Size & 128 \\
				\hline
				VBScript & CPoLS & $B_{VBS,CPoLS}$  & Minibatch Size & 100 \\
				VBScript & CPoLS & $H_{VBS,CPoLS}$  & LSTM Hidden Layer Size & 1500 \\
				VBScript & CPoLS & $E_{VBS,CPoLS}$  & Embedding Layer Size & 128 \\
				VBScript & CPoLS & $W_{VBS,CPoLS}$  & CNN Window Size & 10 \\
				VBScript & CPoLS & $S_{VBS,CPoLS}$  & CNN Window Stride & 5 \\
				VBScript & CPoLS & $F_{VBS,CPoLS}$  & Number of CNN Filters & 128 \\
				\hline
			\end{tabular}
		\end{scriptsize}
	\end{center}
	\caption {Settings for the various model parameters.}
	\label{tab:hyper}
\end{table*}

\noindent
\textbf{End-to-End Learning:}
To train the models described above, we perform an end-to-end learning process.
Since the data available to us is in the form of a sequence and an associated binary label, we need to train the
entire model, solely from this label.
In end-to-end learning, we pass each sequence $B$ through all layers of our model to derive the
probability $p_m$.
Using this probability, with the true label $L \in \{0,1\}$, we measure the cross-entropy loss $\mathcal{L}$.
This loss is used to compute the gradients required for updating the weights in each layer of the model.
Therefore, we simultaneously learn all the parameters for the primary classification objective.

\section{Experimental Results}
\label{sec:eval}
We next evaluate the performance of the proposed neural malware script classifier models on JavaScript and VBScript
files using the data described in Section~\ref{sec:data}. We first start by describing the experimental setup used to generate the results. Instead
of training a single model to detect both JavaScript and VBScript, we train individual models for each script
type since a specific model can better learn to identify the nuances of each particular scripting language. Accordingly, we
first evaluate the LaMP and CPoLS models trained on JavaScript files and then repeat the evaluation for models trained on VBScript files.

\textbf{Experimental Setup:} All the experiments were performed using Keras~(\cite{keras}) with the TensorFlow~(\cite{Tensorflow}) backend. The models were trained and evaluated on a cluster of NVIDIA
K40 graphical processing unit (GPU) cards.
All models were trained with a maximum of 15 epochs, but early stopping was employed if the model
fully converged before reaching the maximum number of epochs.

We did hyperparameter
tuning of the various input parameters for both types of script models, and the results are summarized in
Table~\ref{tab:hyper}. To do so, we first set the other
hyperparameters to fixed values and then vary the hyperparameter under consideration.
For example, to evaluate different minibatch sizes for the JavaScript LaMP classifier, we first set
the LSTM's hidden layer size $H_{JS,LaMP} = 1500$, the embedding dimension to $E_{JS,LaMP} = 128$, the number of LSTM layers $L_{JS,LaMP} = 1$ and
the number of hidden layers in the classifier $C_{JS,LaMP} = 1$. With these settings, we evaluate the classification error rate
on the validation set for the JavaScript dataset.
Table~\ref{tab:hyper} indicates the final hyperparameter settings used for the remainder of the experiments.

\begin{figure*}[tbh]
	\centering
	\begin{subfigure}{.5\textwidth}
		\centering
		\includegraphics[trim = 1.25in 3.0in 1.5in 3.0in,clip,width=1.0\columnwidth]{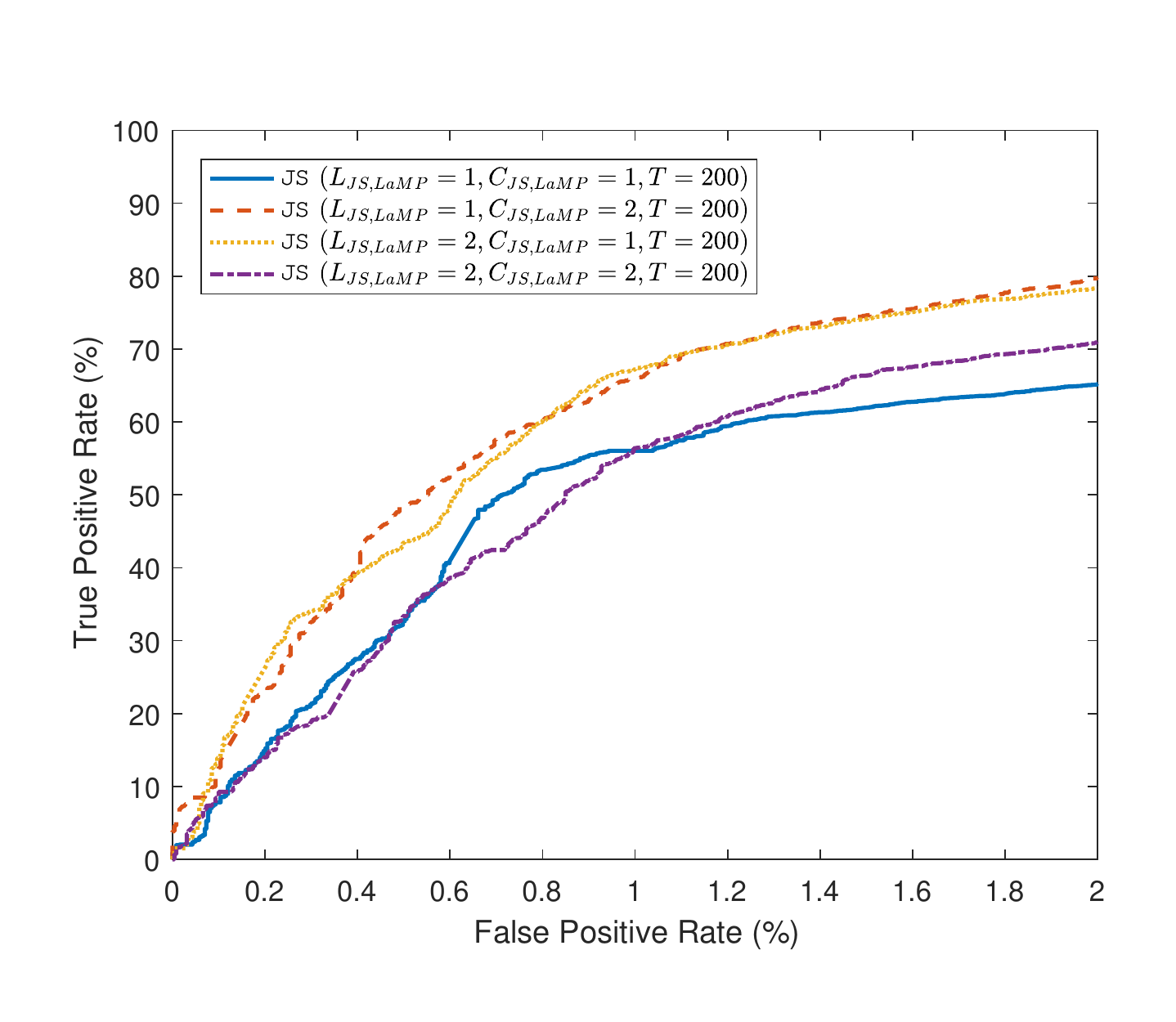}
		\caption{LaMP}
		\label{fig:JS-LaMP-pdf}
	\end{subfigure}%
	\begin{subfigure}{.5\textwidth}
		\centering
		\includegraphics[trim = 1.25in 3.0in 1.5in 3.0in,clip,width=1.0\columnwidth]{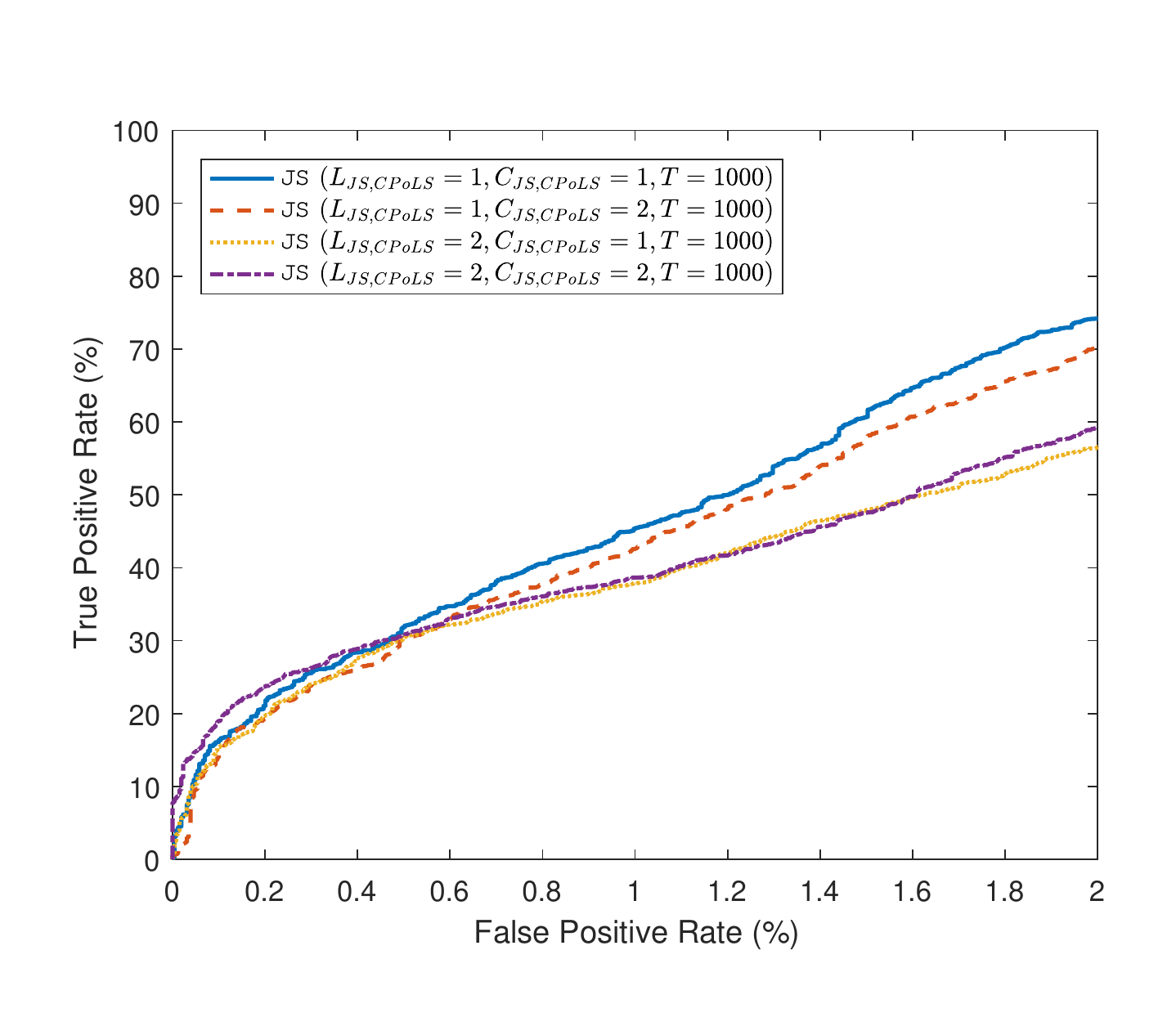}
		\caption{CPoLS}
		\label{fig:JS-CPoLS-pdf}
	\end{subfigure}
	\caption{ROC curves for different JavaScript models.}
	\label{fig:JS-pdf}
\end{figure*}

\textbf{JavaScript:} We evaluate the performance of the LaMP model on the JavaScript dataset in Figure~\ref{fig:JS-LaMP-pdf}
for several different combinations of LSTM stacked layers, $L_{JS,LaMP}$, and classifier hidden layers, $C_{JS,LaMP}$. Similarly,
the CPoLS model is evaluated with the JavaScript files in Figure~\ref{fig:JS-CPoLS-pdf}. For LaMP, adding
either another stacked LSTM layer or classifier hidden layer improves the detection results. On the other hand, the
simplest CPoLS model with one LSTM layer and one neural network hidden layer performs best. For lower FPRs, LaMP
offers significant performance advantages over CPoLS. This result indicates
that sequential modeling of the individual characters in the JavaScript content captures the underlying
behavior compared to a sequential model on the output of the convolutional processing of the subsequences in CPoLS.

At a false positive rate (FPR) of 1\%, the best performing JavaScript LaMP model has a true positive rate of 67.2\% with $L_{JS,LaMP}=2, C_{JS,LaMP}=1$. Similarly for CPoLS with $L_{JS,CPoLS}=1, C_{JS,CPoLS}=1$, the best performing model yields a TPR of 45.3\% at an FPR of 1.0\%

\begin{figure*}[tbh]
	\centering
	\begin{subfigure}{.5\textwidth}
		\centering
		\includegraphics[trim = 1.25in 3.0in 1.5in 3.0in,clip,width=1.0\columnwidth]{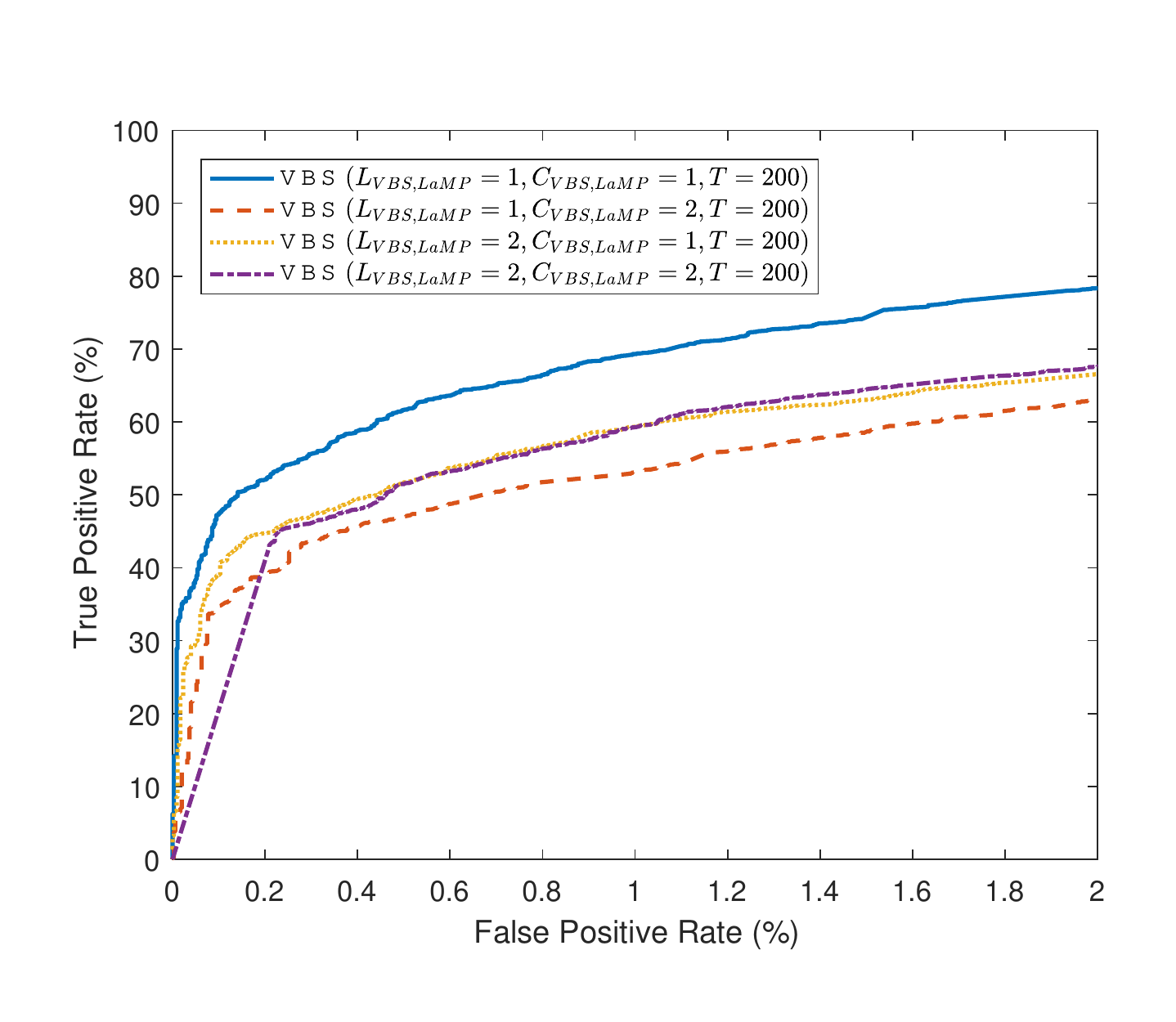}
		\caption{LaMP}
		\label{fig:VBS-LaMP-pdf}
	\end{subfigure}%
	\begin{subfigure}{.5\textwidth}
		\centering
		\includegraphics[trim = 1.25in 3.0in 1.5in 3.0in,clip,width=1.0\columnwidth]{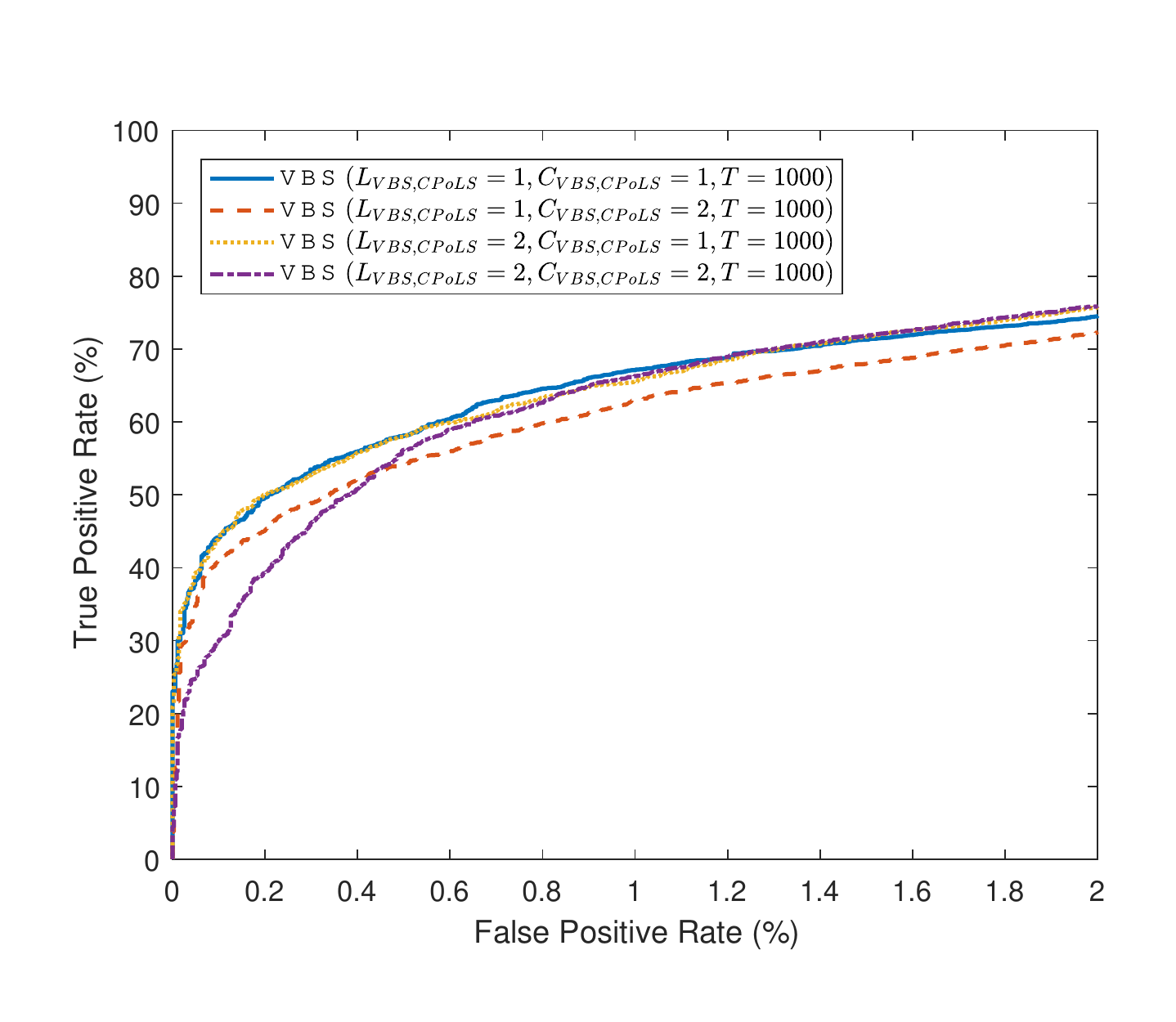}
		\caption{CPoLS}
		\label{fig:VBS-CPoLS-pdf}
	\end{subfigure}
	\caption{ROC curves for different VBScript models.}
	\label{fig:VBS-pdf}
\end{figure*}

\textbf{VBScript:} Next we evaluate the LaMP and CPoLS models for VBScript in Figures~\ref{fig:VBS-LaMP-pdf} and~Figure~\ref{fig:VBS-CPoLS-pdf}, respectively. Similar to the JavaScript CPoLS model results, the simplest LaMP and CPoLS VBScript models with
a single LSTM layer and classifier hidden layer offer the best, or nearly the best, performance compared to the more complex models.
At an FPR of 1.0\%, the TPR for the LaMP model is 69.3\% with $L_{VBS,LaMP}=1, C_{VBS,LaMP}=1$. Similarly, CPoLS yields a
TPR of 67.1\% with $L_{VBS,CPoLS}=1, C_{VBS,CPoLS}=1$ at this FPR = 1.0\%.

\section{Discussion}
\label{sec:discussion}
In this section, we consider several limitations of the proposed \Sys neural malware script classification system. These include
limitations due to the size of the GPU memory 
and adversarial learning-based attacks.

One limitation is the maximum sequence length, $T = 200$, employed by the LaMP models. This parameter value
was primarily chosen because it allows the LaMP models to be trained in the 12 GB of SDRAM on the NVIDIA K40. If the
length was increased much beyond this value, we could not train all the models investigated in
this study. It may be possible that more advanced GPUs that are released in the future, and contain more GPU memory, might allow better
performance if the maximum sequence length can be extended.

Attacks based on adversarial learning are another important concern.
Both architectures used in this study include recurrent LSTM and possibly deep
neural network (DNN) components. While researcher have not directly attacked LSTM structures using adversarial learning-based attacks,
~\cite{PapernotRnn2016} have shown that standard RNN cells (\textit{i.e.}, SimpleRNN) are vulnerable by unrolling the recurrent loop.
Like DNNs, this unrolled structure can then be attacked using a number of methods for crafting adversarial samples~\cite{HuAdversarialMalwareGan,papernot2015limitations}.
One possible defense is to run the classifier in a secure enclave such as Intel's SGX~(\cite{ohrimenko2016oblivious}). Other
defenses including distillation and ensembles have been explored for PE files~(\cite{Grosse2017a,Stokes2017Adversarial}).

\section{Related Work}
\label{sec:related}
%
\textbf{JavaScript:}
%
~\cite{Maiorca15} propose a static analysis-based system to detect malicious
PDF files which use features constructed from both the content of the PDF, including JavaScript, as well as its structure. Once
these features are extracted, the authors use a boosted decision tree trained with the AdaBoost algorithm to detect malicious PDFs.
%
%
%
~\cite{Cova2010} use the approach of anomaly detection for detecting malicious JavaScript code.
They learn a model for representing normal (benign) JavaScript code, and then use it during the detection of anomalous code.
They also present the learning of specific features that helps characterize intrinsic events of a drive-by download.
%
~\cite{Hallaraker2005} present an auditing system in Mozilla  for JavaScript interpreters.
They provide logging and monitoring on downloaded JavaScript, which can be integrated with
intrusion detection systems for malicious behavior detection.
In~\cite{Likarish2009}, they classify obfuscated malicious JavaScript using several different types of classifiers including Naive Bayes, an Alternating Decision Tree (ADTree), a Support Vector Machine (SVM) with using the Radial Basis Function (RBF) kernel, and the rule-based Ripper algorithm. In their static analysis-based study, the SVM performed best based on tokenized unigrams and bigrams chosen by feature selection.
A PDF classifier proposed by~\cite{Laskov2011} uses a one-class SVM to
detect malicious PDFs which contain JavaScript code. Laskov's system is based solely on static
analysis. The features are derived from lexical analysis of JavaScript code extracted from the PDF
files in their dataset.
~\cite{Corona2014}, propose Lux0R, a system to select API references for the detection of malicious JavaScript in PDF documents. These
references include JavaScript APIs as well as functions, methods, keywords, and constants. The authors propose a discriminant analysis feature selection method.
The features are then classified with an SVM, a Decision Tree and a Random Forest model.
Like ScriptNet, Lux0R performs both static and dynamic analysis. However, they do not use deep
learning and require the extraction of the JavaScript API references.
~\cite{Wang2016} use deep learning models in combination with sparse random projections, and logistic regression.
They also present feature extraction from JavaScript code using auto-encoders.
While they use deep learning models, the feature extraction and model architectures limit the information extractability from JavaScript code.
%
~\cite{Shah2016} propose using a statistical n-gram language model
to detect malicious JavaScript. Our proposed system uses an LSTM neural model for the language model instead of the n-gram model proposed by~\cite{Shah2016}.
Other papers which investigate the detection of malicious JavaScript include~\cite{Liu2014,Schutt2012,Wang2013,Xu2012,Xu2013}.

\textbf{VBScript:}
While more research has been devoted to detecting malicious JavaScript, partly because of
its inclusion in malicious PDFs, only a few previous studies have considered malicious VBScript.
In~\cite{Kim2006}, a conceptual graph is first computed for VBScript files, and new malware
is detected by identifying graphs which are similar to those of known malicious VBScript files. The
method is based on static analysis of the VBScripts.
%
~\cite{Wael2017} propose a number of different classifiers to detect malicious VBScript including Logistic Regression, a Support Vector Machine with an RBF kernel, a Random Forest, a Multilayer Perceptron, and a Decision Table. The features are created based on static analysis. The best performing classifier in their study is the SVM. In~\cite{Zhao2010}, they detect malicious applets, JavaScript and VBScript based on a method which models immunoglobulin secretion.

\textbf{Other File Types:} A number of deep learning models have been proposed for detecting malicious PE files including~\cite{BenMalware,Dahl2013,Huang2016,Kolosnjaji,PascanuMalware}.
In particular, a character-level CNN has been proposed for detecting malicious PE files~(\cite{BenMalware}) and Powershell script files~(\cite{Hendler2018}).
~\cite{Raff2017} discuss a model which is similar to CPoLS but noted it did not work for PE files. They did not provide any results
for their model.

\section{Conclusions}
\label{sec:conc}
Malicious script classification is an important
problem facing anti-virus companies. Failure to
detect a malicious script may result in a
successful spearphishing, ransomware, or drive-by
download attack. Neural language models have shown promising
results in the detection of malicious executable files.
Similarly, we show that these types of models
can also detect malicious JavaScript and VBScript files with relatively high true
positive rates at low false positive rates. These results
are even more remarkable because the best performing models only
utilize the first 200 characters in the script, making them
fast for large-scale production.

The performance results confirm that the \MPL and \SCL architectures
using LSTM and CNN neural models
are able to learn and generate representations of byte sequences
in the scripts.
In particular, the LaMP JavaScript malware script classification model using two
LSTM layers and one dense neural network layer offers the best results,
while for VBScript malware, the LaMP model with one LSTM and one hidden layer
is significantly better than the competing models.
The embeddings generated in these models, therefore, capture important
sequential information from within the script file and help to predict their
malicious nature through neural training over these embeddings.


\vskip 0.2in
\bibliography{Esorics2018}

\end{document}